\documentclass[prl,twocolumn,aps]{revtex4-1}

\usepackage{amssymb,graphicx}
\usepackage{amsmath,bm}
\usepackage{color}
\begin{document}

\title{Instantons from Perturbation Theory}
\author{Marco~Serone$^{a,b}$, Gabriele~Spada$^a$, and Giovanni~Villadoro$^{b}$}

\affiliation{
 \sl $^{a}$ 
SISSA International School for Advanced Studies and INFN Trieste, 
Via Bonomea 265, 34136, Trieste, Italy \\
 \sl $^{b}$ Abdus Salam International Centre for Theoretical Physics, 
Strada Costiera 11, 34151, Trieste, Italy
}
\date{\today}
%
%\vspace*{0.1in}
%
\begin{abstract}
In quantum mechanics and quantum field theory perturbation theory 
generically 
requires the inclusion of extra contributions non-perturbative in the coupling, 
such as instantons, to reproduce exact results. 
We show how full non-perturbative results can be encoded in a suitable modified perturbative series
in a class of quantum mechanical problems. We illustrate this explicitly in examples which are known to 
contain non-perturbative effects, such as the (supersymmetric) double-well 
potential, the pure anharmonic oscillator, and the perturbative expansion 
around a false vacuum. 
\end{abstract}
%
%\today
\maketitle

The coefficients of saddle-point expansions of (path) integrals generically
grow factorially and produce  non-convergent asymptotic series. 
In special cases exact results can be obtained from such series by  ``Borel resummation'',
consisting in taking the Laplace transform of the function obtained by resumming
the original series after dividing their terms by a factorially growing coefficient.
However this procedure generically fails, because exact results take the form of trans-series, 
i.e. series in powers of the coupling $\lambda$, $e^{-1/\lambda}$ and $\log(-\lambda)$, which capture behaviours non-perturbative in the coupling $\lambda$.

There are several well understood tools which determine the properties of perturbative series 
of finite-dimensional integrals. For path integrals some results are available in specific cases 
(mostly within quantum mechanics or supersymmetric theories using localization
techniques) but a systematic characterization of the behavior of their perturbation theories is still lacking. 
Quantum mechanics (QM) represents the simplest playground where to test our understanding 
of the interplay between perturbative and non-perturbative effects in path integrals, 
for exact results can be extracted efficiently by other means. 
In this letter we present a method to recover the full non-perturbative answer 
from a deformation of the perturbative series in a certain class of QM problems that contain 
non-perturbative effects. Such technique could be used to improve numerical computations
based on perturbation theory in QM and might, in principle, be extended
to higher dimensional quantum field theories (QFT).

Consider one-dimensional quantum mechanical systems described by the Hamiltonian
\begin{equation}
H=\frac{p^2}{2}+V(x;\lambda)\,, 
\end{equation}
where the potential $V(x;\lambda)$ is a regular function of $x$ describing a bounded system (i.e. $\lim_{|x|\to \infty}V(x)=\infty$). If the dependence on the coupling $\lambda$ is such that $V(x;\lambda)=V(x\sqrt{\lambda};1)/\lambda$
then the perturbative expansion in $\lambda$ coincides with the $\hbar$ expansion. We call such potential
\emph{classical}. If $V_0(x; \lambda)$ and $V_1(x;\lambda)$ are two such potentials then the combination
$V(x;\lambda)=V_0(x;\lambda)+\lambda V_1(x;\lambda)$ is a sum of a classical contribution ($V_0$) and a quantum one $(V_1)$.

Consider now the classical anharmonic oscillator
\begin{equation} \label{eq:Vao}
V^{ao}(x;\lambda)=\frac{1}{2} x^2+\frac{\lambda}{2}x^4\,,
\end{equation}
whose energy eigenvalues at small $\lambda$ are close to those of the harmonic oscillator 
 $E^{ao}_n=n+\frac12+{\cal O}(\lambda)$. 
By studying the analytic properties of the eigenvalues,
their perturbative series has been shown to be Borel resummable to the exact result 
\cite{Simon:1970mc,Graffi:1990pe}. 
 The conditions under which Borel resummability holds  
 can be characterized more systematically \cite{longpaper}, 
 extending the result to other potentials and observables.
Similar results seem to hold also in higher dimensions, 
i.e. $\lambda\phi^4$ theories in two and three dimensions 
\cite{Eckmann,Magnen}. 

Note that the Borel resummability of the theory crucially depends on the presence
and sign of the quadratic term $x^2$. In fact, in the case of the double-well potential,
where the quadratic term is negative, perturbation theory in $\lambda$ around the minima 
is known to be non-Borel resummable. Non perturbative effects, instantons, are needed
to cure the ambiguities of the Borel transform and to reproduce the exact answer.
A particularly interesting example is the supersymmetric double-well potential where 
perturbative corrections to the vacuum energy vanish at all orders, making the perturbative series
trivially Borel resummable. Still supersymmetry is broken non perturbatively~\cite{Witten:1981nf} 
and the vacuum energy is lifted by instanton contributions.

For the special case of vanishing mass term, 
i.e. for the \emph{pure} anharmonic oscillator
\begin{equation}
V^{pa}(x)=\frac12 x^4\,,
\end{equation}
the situation is more subtle. This case corresponds to the strong coupling limit, $\lambda\to\infty$, of the 
anharmonic potential, obtained after the rescaling $x\to x/\lambda^{1/6}$, $p\to p \lambda^{1/6}$,
and $H\to H/\lambda^{1/3}$. As such, normal perturbation theory (i.e. perturbation theory around the harmonic oscillator) cannot be used. In this special case a different semi-classical
expansion in $\hbar$, the WKB approximation, is more useful. 
Not only such perturbative expansion is not Borel resummable 
but real instantons are not enough to reproduce the full answer 
\cite{BPV}---complex saddles are also required  to recover the full 
result (see \cite{Grassi:2014cla} for a recent reanalysis).

In this work we present a different approach, which instead uses the normal 
perturbation theory and exploits the Borel resummability of the anharmonic potential, to show that the full result can be 
recovered by a single perturbative series (with no need of trans-series) in all the cases above, even at strong coupling ($\lambda\gtrsim 1$) \cite{Caswell:1979qh}.

We will make use of the following property. If $V_0(x;\lambda)$ is a classical potential
with a perturbation theory which is Borel resummable to the exact result 
(as in the case of the anharmonic oscillator), 
the perturbative series of a QM system with potential $V(x;\lambda)=V_0(x;\lambda)+\lambda V_1(x;\lambda)$ 
is still Borel resummable to the exact result for any $V_1$ if 
$\lim_{|x|\to \infty} V_1(x;1)/V_0(x;1)=0$. 
We will provide a derivation of this
property in \cite{longpaper} using Leftschetz-thimbles techniques along the lines of \cite{Witten:2010cx}, 
in the current work we limit ourselves to provide numerical evidence 
in a number of non-trivial examples.

The property above allows us to compute a certain class of QM problems through a Borel
resummable perturbative expansion even if the initial naive $\hbar$ expansion was not.
Imagine we have a system with potential $V(x;\lambda)$ and we want to compute a quantity
at some value of the coupling $\lambda=\lambda_0$. If we can find a potential ${\hat V}(x;\lambda)=V_0(x;\lambda)
+\lambda V_1(x;\lambda)$ satisfying the conditions above and such that ${\hat V}(x;\lambda_0)=V(x;\lambda_0)$
we will have a way to extract the full answer by computing the perturbative
expansion in $\lambda$ of the modified potential $\hat V(x;\lambda)$ and setting
$\lambda=\lambda_0$ after the Borel resummation. We name such
expansion ``exact perturbation theory'' (EPT), while we refer to the trans-series 
originating from the potential $V(x;\lambda)$ as ``standard perturbation theory'' (SPT).

In practical computations where the perturbative series is truncated at a fixed order,
the accuracy of EPT depends on the choice of the potential $\hat V$ and on the value of the coupling constant.

As first application consider the anharmonic potential in eq.~(\ref{eq:Vao}) perturbed at the quantum level
by a negative mass term, namely
\begin{equation}
 {\hat V}^{pa}(x;\lambda)=\left[\frac12 x^2+ \frac{\lambda}2 x^4\right] -\lambda\left[ \frac{1}{2} x^2\right]\,.
 \end{equation} 
At $\lambda=1$  the potential  ${\hat V}^{pa}(x;1)$ reduces to the potential 
$V^{pa}(x)$ of the pure anharmonic oscillator. According to our criterion
a Borel resummation of the perturbative series of ${\hat V}^{pa}(x;\lambda)$ computed at $\lambda=1$ 
should therefore reproduce the exact results of the pure anharmonic potential.
Perturbative coefficients of eigenvalues and eigenfunctions in QM can be efficiently  
computed using the recursion relations by Bender and Wu \cite{Bender:1990pd,Sulejmanpasic:2016fwr}.

Using only the first ten orders of perturbation theory (and performing a Pad\'e-Borel resummation) 
the estimate for the ground state energy $E_0^{pa}=0.5302$ matches the accuracy obtained by the
WKB expansion of \cite{Grassi:2014cla}, using 320 orders of perturbative expansions and including the leading non-perturbative complex and real saddle contributions.
The accuracy can easily be improved by including higher-order terms without
the need of extra non-perturbative effects;  for example with 320 orders of perturbation theory 
the accuracy reaches the $10^{-46}$ level. The result agrees with that obtained 
using Rayleigh-Ritz methods (see e.g.~\cite{vieira}), 
which remains however the most efficient for this type of problems.

Few remarks are in order. The pure anharmonic potential is a genuinely non-perturbative problem,
in the sense that there is no small parameter to expand around. The modified potential $\hat V(x,\lambda)$
admits a perturbative expansion but it coincides with the pure anharmonic potential only at strong coupling
$\lambda=1$. Still the Borel resummability of the theory allows us to reconstruct the full answer from the
perturbative terms alone without having to include non-perturbative effects (unlike the WKB expansion). 
The precision that can be achieved in this way is remarkable, especially considering that we are working at $\lambda=1$. We should say that the WKB approach is particularly weak for the ground state while it improves
considerably for higher energy states, still the expansion is not Borel resummable and the full answer requires the inclusion of non-trivial saddles. EPT instead does not require extra non-perturbative contributions and the full answer is always contained in the perturbative series. We checked
numerically also the wave-function as well as higher excited states, which can be computed 
with similar high precision (although the accuracy slowly reduces for higher energy states), 
confirming that indeed all observables are Borel-resummable to the exact result.  
We also checked numerically other pure anharmonic potentials of the type $V(x)=x^{2n}$ with $n>2$.

We now turn to the discussion of the ground state eigenvalue of the supersymmetric double-well
or Fokker-Planck potential
\begin{equation} \label{eq:susypot}
V^{sdw}(x;\lambda)=\frac{\lambda}2 \left[x^2-\frac{1}{4\lambda}\right]^2+\sqrt{\lambda}x\,,
\end{equation}
which is the sum of a classical symmetric double-well potential (first term) 
plus a quantum linear tilt correction (second term). 
The latter can be interpreted as the contribution to the classical bosonic
potential from integrating out the fermionic variables. Because of supersymmetry 
the ground state energy vanishes at all orders in $\lambda$. Supersymmetry is however 
known to be broken non-perturbatively~\cite{Witten:1981nf}.
At weak coupling the leading contribution comes from instantons interpolating 
the two classical vacua at $x=\pm 1/\sqrt{4\lambda}$. This contribution has been 
computed at high orders in \cite{Jentschura:2004jg} using a generalized Bohr-Sommerfeld 
quantization formula \cite{ZinnJustin:1983nr} and the result agrees extremely well at weak 
coupling with the exact answer 
from non-perturbative numerical computations such as Raylegh-Ritz methods. 
While the instanton computation is particularly efficient 
at weak coupling it becomes useless at strong coupling---for $\lambda\gtrsim 1$ 
the perturbative expansion around the leading instanton 
solution is divergent and non-Borel resummable, 
an infinite series of multi-instanton solutions and the corresponding perturbative series 
have to be properly resummed altogether, making the full approach impractical.
The presence of such instantons may seem puzzling since the full
quantum corrected potential (\ref{eq:susypot}) has no non-trivial 
finite-action solutions. 
This fact led several authors \cite{Balitsky:1985in,Behtash:2015zha} 
to reinterpret the non-perturbative effects in terms of complex instantons. 
From our point of view however there is no puzzle: The asymptotic properties of the $\lambda$ expansion 
are determined only by the classical part of eq.~(\ref{eq:susypot}), which is a symmetric double-well potential known to have a non-Borel resummable perturbative expansion 
and requiring the inclusion of  the corresponding (real) instantons.
For the ground state energy the perturbative saddle-point expansion around the minimum of the full
quantum corrected potential (\ref{eq:susypot}) is instead Borel resummable 
to the exact result \cite{longpaper}. 

Consider now the potential
\begin{align} \label{eq:Vsdwh}
\hat V^{sdw}(x;\lambda,\lambda_0)=&\left[ \frac{1}{32\lambda}+\frac{\lambda_0}{2}x^2+\frac{\lambda}{2}x^4\right] \nonumber \\
&+\sqrt{\lambda}x - \lambda \left[1+\frac1{2\lambda_0} \right]\frac{x^2}2\,,
\end{align}
which for $\lambda_0=\lambda$ reduces to eq.~(\ref{eq:susypot}), i.e. $\hat V^{sdw}(x;\lambda,\lambda)=V^{sdw}(x;\lambda)$.
Expanding in $\lambda$ at fixed $\lambda_0$ the terms in the first line of eq.~(\ref{eq:Vsdwh}) represent the
classical part of the potential and correspond to a Borel-resummable anharmonic oscillator.
Similarly to the case of the pure anharmonic oscillator we conclude that the perturbative 
expansion in $\lambda$ (at fixed $\lambda_0$) is Borel resummable to the exact result for any $\lambda$. 
Setting $\lambda=\lambda_0$ we thus recover the result for the supersymmetric double-well. 

Note that the choice of $\hat V^{sdw}(x;\lambda,\lambda_0)$ above
is particularly good at moderate and large values of $\lambda_0$ but is not ideal for small 
values of the coupling.
In the latter regime, non-perturbative effects become exponentially small
and anyway harder to be resolved, as discussed below.

\begin{figure}[t!]
\centering
\includegraphics[width=0.48\textwidth]{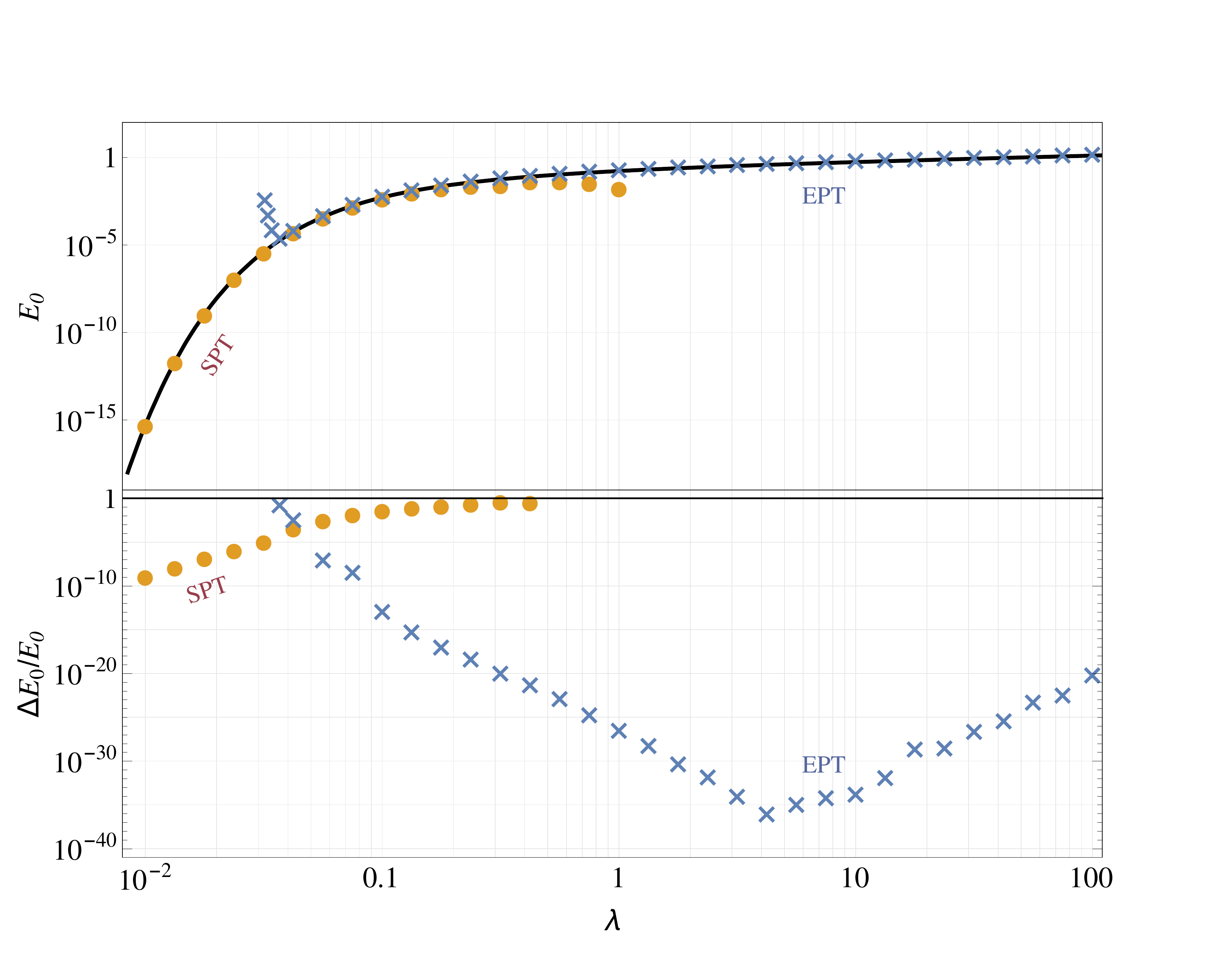}
\caption{\label{fig:susy} The ground state energy (top) and the relative error (bottom)
as a function of the coupling $\lambda$ for the supersymmetric double-well 
potential (\ref{eq:susypot}) computed using the SPT up to the 9th 
order from \cite{Jentschura:2004jg} (orange dots) and using our EPT (blue crosses). 
The exact result (black line) has been computed 
via a Rayleigh-Ritz method.
}
\end{figure}

In fig.~\ref{fig:susy} we compare the accuracy of SPT from
 the leading instanton contribution of $V^{sdw}(x;\lambda)$ computed in \cite{Jentschura:2004jg} 
 against EPT using 200 orders and Pad\'e-Borel resummation. 
At small coupling SPT becomes more accurate, 
since the leading order instanton formula provides the asymptotic $\lambda\to 0$ limit of the result.
On the other hand EPT is less and less accurate at weak coupling---such expansion 
is not supersymmetric,  at $\lambda=\lambda_0$
perturbative corrections at each order cancel up to the next order, leading to an
asymptotic series whose Borel-resummation encodes the full non-perturbative result.
This means that at small coupling more orders of perturbation theory are required
to capture the exponentially small non-perturbative effects, with a corresponding
weakening of accuracy. Notice however that there is a non-trivial interval of small couplings 
where both SPT and EPT are under control 
and agree at the $10^{-4}$ level. At strong coupling on the other hand
 EPT works extremely well. Paradoxically,
while the instanton calculation only works
in the perturbative limit, EPT works best in the non-perturbative regime!

We can repeat a similar analysis for the non supersymmetric double-well potential
\begin{equation} \label{eq:potdw}
V^{dw}(x;\lambda)=\frac{\lambda}2 \left[x^2-\frac{1}{4\lambda}\right]^2\,.
\end{equation}
In this case the ground state energy receives corrections also from the perturbative expansion
in $\lambda$, which is not Borel resummable. The combination of the perturbative corrections and
the instanton ones requires some care, see e.g. \cite{Jentschura:2001kc}. As usual this approach
works better and better at weak coupling while it breaks down at strong coupling. As before we
can recover the same result by considering, for example, the potential
\begin{equation}
\hat V^{dw}(x;\lambda,\lambda_0)=\left[\frac{1}{32\lambda}+\frac{\lambda_0}{2}x^2+\frac{\lambda}{2}x^4\right]
- \lambda\left[1+\frac1{2\lambda_0} \right]\frac{x^2}2,
\end{equation}
whose classical part (in the first brackets) corresponds to a Borel resummable anharmonic oscillator
and it matches the potential (\ref{eq:potdw}) for $\lambda=\lambda_0$. Using 300 orders of perturbation theory at $\lambda=\lambda_0=0.03$ we get a ground state energy $E_0^{dw}=0.4546(15)$, which differs from the exact value ($0.4531$) only by $1.5\cdot 10^{-3}$. 
On the other hand, the difference between the two lowest energies,
which represents the size of the instanton contribution and the level of precision that can be reached
by just truncating the perturbative asymptotic series, is $0.02$, i.e. one order of magnited larger. 
The accuracy rapidly improves increasing the coupling. 
Already at $\lambda=0.04$, the difference between EPT and
the exact answer drops to $3\cdot 10^{-9}$, while the instanton contributions amount to $0.06$. 
It is thus clear that EPT is able to
correctly reproduce the full answer, automatically combining perturbative and multi-instanton 
contributions  at weak coupling. Considerations similar to those  of the supersymmetric 
double-well apply regarding the complementarity of SPT
versus EPT: while the former is more powerful at weak coupling the latter 
extends also at strong coupling. Furthermore note that %, even in the  instanton based computation, 
observables that receive both a perturbative and a non-perturbative contribution 
in SPT (as the ground state energy) require a non-trivial resummation before 
the instanton contributions could improve the accuracy.

The final example we present, although very similar to the previous ones,
is maybe the most amusing from a theoretical point of view.
Consider the potential
\begin{equation} \label{eq:Vfv}
V^{fv}(x;\lambda)= \frac12 x^2-\frac32 \sqrt{\lambda}\, x^3 +\frac{\lambda}2 x^4 \,,
\end{equation}
corresponding to an asymmetric double-well with a false vacuum at $x=0$ and a true one at 
$x=2/\sqrt\lambda$.
Perturbation theory around the false vacuum is known to be non-Borel resummable, with the
corresponding ambiguity related to the instanton describing the tunneling to the true vacuum.
Information about the true vacuum is completely non-perturbative from the point of view of the
perturbative expansion around the false vacuum. Consider now the potential
\begin{equation} \label{eq:Vfvh}
\hat V^{fv}(x;\lambda,\lambda_0)=\left[\frac12 x^2+\frac{\lambda}2 x^4 \right]-\lambda\left[\frac32 \frac{\sqrt{\lambda}}{\lambda_0}\, x^3\right]\,,
\end{equation}
whose classical part is a Borel resummable anharmonic oscillator around the false vacuum. 
At small coupling the potential is an asymmetric anharmonic oscillator with a unique vacuum at $x=0$. 
As $\lambda$ increases a new vacuum develops becoming deeper than the one at the origin. 
At $\lambda=\lambda_0$ the potential coincides with the one in eq.~(\ref{eq:Vfv}). 
Analogously to the previous cases we therefore expect that the perturbative $\lambda$
expansion around the origin of $\hat V^{fv}(x;\lambda,\lambda_0)$ reproduces 
the exact results of the potential $V^{fv}(x;\lambda)$ for $\lambda=\lambda_0$ 
when the perturbative series is properly Borel resummed.

At $\lambda=1$ the ground state energy is $E_0^{fv}=-0.828$, deeply below zero. 
Using $\hat V^{fv}(x;\lambda,1)$ with 280 orders of perturbation theory  and performing a 
Pad\'e-Borel resummation we get the estimate $\hat E_0^{fv}=-0.847$. This result is in quite
good agreement with the exact value, especially considering that 
the ground state energy of the corresponding classical anharmonic potential 
around the false vacuum is large and positive (+0.7). 
Despite in this example the accuracy is not on par with the previous cases,
it is notable that perturbation theory around the false vacuum can still be used to recover
a completely non-perturbative result, such as the energy of the true vacuum, with a few percent accuracy.

The examples above  show how, in a class of QM problems, a suitable modification 
of the perturbative expansion allows us to recover all non-perturbative effects
from perturbation theory alone, in the spirit of the resurgence program \cite{ecalle},
but without the need of trans-series. The high accuracy (especially at strong coupling) 
of the numerical checks presented above confirms the absence of extra 
contributions not accounted for by the EPT.
The success of this expansion can be understood more easily by extending Lefschetz-thimble 
techniques to the QM path integral \cite{longpaper}, suggesting that similar methods might
also apply to more general theories. 
Complications such as ultraviolet divergences and the possible presence 
of phase transitions could make the extension to QFT less straightforward.
However a certain class of non-trivial QFT have been proven to be Borel resummable 
\cite{Eckmann,Magnen} and may serve as starting point (the analogue of $V_0$)
to apply the technique to more general theories.
The idea of EPT may then represent a new avenue for the study of 
strongly coupled theories, by allowing all non-perturbative effects 
to be recovered from perturbation theory alone.

%\section*{Acknowledgments}
We thank J.~Elias-Mir\'o, E.~Poppitz, T.~Sulejmanpasic for discussions.
This work is supported in part by the ERC Advanced Grant no.267985 (DaMeSyFla).

\end{document}